\documentclass[final,5p,times,twocolumn]{elsarticle}

\makeatletter
\def\@author#1{\g@addto@macro\elsauthors{\normalsize%
    \def\baselinestretch{1}%
    \upshape\authorsep#1\unskip\textsuperscript{%
      \ifx\@fnmark\@empty\else\unskip\sep\@fnmark\let\sep=,\fi
      \ifx\@corref\@empty\else\unskip\sep\@corref\let\sep=,\fi
      }%
    \def\authorsep{\unskip,\space}%
    \global\let\@fnmark\@empty
    \global\let\@corref\@empty  
    \global\let\sep\@empty}%
    \@eadauthor={#1}
}
\def\ps@pprintTitle{%
 \let\@oddhead\@empty
 \let\@evenhead\@empty
 \def\@oddfoot{}%
 \let\@evenfoot\@oddfoot}
\makeatother

\usepackage{graphicx}
\usepackage{amssymb}
\usepackage{amsthm}
\usepackage{amsmath}
\usepackage{float}
\usepackage{subcaption}
\usepackage{comment}
\usepackage{color}
\usepackage{multirow}
\usepackage{array}
\usepackage{bm}
\usepackage{url}

\newcommand{\p}{\hat{p}}
\newcommand{\bp}{\mathbf{p}}
\newcommand{\bQ}{\mathbf{Q}}

\newcommand{\q}{\tilde{q}}
\newcommand{\E}{\mathbb{E}}
\newcommand{\Cov}{\mathrm{Cov}}
\newcommand{\Var}{\mathrm{Var}}

\usepackage{verbatim}

\usepackage[dvipsnames]{xcolor}

\usepackage[colorinlistoftodos]{todonotes}

\usepackage[normalem]{ulem} 

\usepackage{lineno}

\biboptions{comma,round,authoryear}

\begin{document}

\begin{frontmatter}


\title{An information-theoretic approach to the analysis of location and co-location patterns}


\author{Alje van Dam \fnref{label1,label2}\corref{cor1}}
\ead{A.vanDam@uu.nl}

\author{Andres Gomez-Lievano \fnref{label3}}

\author{Frank Neffke \fnref{label3}}

\author{Koen Frenken \fnref{label1}}


\cortext[cor1]{Corresponding author}

\fntext[label1]{Copernicus Institute of Sustainable Development, Utrecht University}

\fntext[label2]{Center for Complex Systems Studies, Utrecht University}

\fntext[label3]{Growth Lab, Center for International Development, Harvard University}



\begin{abstract}
We propose a statistical framework to quantify location and co-location associations of economic activities using information-theoretic measures. We relate the resulting measures to existing measures of revealed comparative advantage, localization and specialization and show that they can all be seen as part of the same framework. Using a Bayesian approach, we provide measures of uncertainty of the estimated quantities. Furthermore, the information-theoretic approach can be readily extended to move beyond pairwise co-locations and instead capture multivariate associations. To illustrate the framework, we apply our measures to the co-location of occupations in US cities, showing the associations between different groups of occupations. 


\end{abstract}

\begin{keyword}
pointwise mutual information \sep Kullback-Leibler divergence \sep revealed comparative advantage (RCA) \sep specialization \sep localization \sep co-agglomeration 



\end{keyword}

\end{frontmatter}

\newpage










\section{Introduction}





The recognition of differential specialization patterns lies at the heart of economics since the works of Adam Smith and David Ricardo. Economists studying task assignments \citep{Roy1951, Sattinger1993}, urban economies \citep{Ellison1997, Ellison2010}, or international trade \citep{Balassa1965, Krugman1991b}, all stress the fact that different economic entities specialize in different activities. Scholars in each of these fields have relied on indices that quantify, for example, the revealed comparative advantage of exports, the specialization of regions, and the extent of localization and (co-)agglomeration of industries. However, these indices are often used ad hoc and lack a clear statistical foundation. In this paper, we propose a statistical framework from which such measures can be derived. Although the methodology generalizes immediately to other contexts, to fix ideas, we focus on economic geography and derive measures of (co-)location, specialization and localization from a single statistical framework, revealing the internal connections between these concepts.

We treat (co-)location as the realizations of two categorical random variables: the location and the type of an economic activity. We use the Pointwise Mutual Information (PMI) to express the association between a location and the type of an activity in terms of the information that the type of a unit of activity (e.g. a person's occupation) gives about the unit's location (e.g. the city where that person works). Next, we show how the PMI can be used to quantify the association between two activity types in terms of how much information observing a particular activity type in a location gives about observing another activity type in the same location. That is: if we observe a pair of people from the same city, how much information does the occupation of one of them provide about the likely occupation of the other?

The information-theoretic basis that underlies the PMI ensures that the framework is explicit about the null models, priors and data-generating processes we assume. This puts the measurement of location and co-location on a rigorous statistical footing. Furthermore, we show how the PMI can be estimated from data on the counts of activities across locations. To do so, we use a Bayesian framework that assumes that the data on the presence of units of economic activities across locations are generated from a multinomial distribution. This Bayesian estimation framework resolves some well-known measurement issues and provides a measure of uncertainty for the estimated quantities.

Metrics based on Information Theory such as the PMI have found various applications in economics \citep{Theil1967}, and are uniquely derived from axioms about how information can be gained from probability distributions \citep{Shannon1948, Cover2005}. One of their key properties is that they can be aggregated and decomposed to form well-defined measures that have an interpretation in terms of information, by taking expectations. This allows the use of the PMI as a building block of information-theoretic measures that describe properties at the location, activity, or even system level. 


We show how the resulting measures can be related to well-known existing indices of localization and specialization. In particular, at the level of location-activity pairs -- as exemplified in country-product or city-industry data -- our metric of association, the PMI, is conceptually similar to the logarithm of the widely used index of revealed comparative advantage (RCA) \citep{Balassa1965}.\footnote{The RCA is also known as the Location Quotient in the regional science literature \citep{Isard1960}. } This provides an information-theoretic motivation for considering the logarithm of the RCA index, which has the practical advantage that it overcomes the RCA index's problem of distributional skew. Moreover, the Bayesian estimation procedure ensures that the measure always attains finite values, and suggests a natural measure of uncertainty for the estimates. 

Building on the location-activity PMI, we can furthermore derive a measure for the localization of economic activities, that is, for the degree to which economic activities are spatially constrained. We do so by calculating an activity's expected PMI (i.e., the expected association of the activity with a given location) over all locations. This yields the Kullback-Leibler divergence, which has been proposed as a measure of localization before \citep{Mori2005}. Likewise, we can calculate the expected location-activity PMI of a particular location across all activity types. This average association of a location with given activities provides a measure of specialization that is conceptually similar to Krugman's specialization index \citep{Krugman1991b}.

Finally, we apply the PMI to the distribution of co-located pairs of economic activity, which gives the probabilities that pairs of activities are located in the same geographic unit. This provides a measure of spatial association between economic activities. Such measures may reveal positive or negative co-location forces, and are conceptually similar to widely used (co-)agglomeration measures \citep{Ellison1997, Ellison2010}. Here we derive such measures from first principles, which clarifies their underlying assumptions and statistical properties.

As in the case of location-activity pairs, marginalizing the PMIs of co-located activity-activity pairs yields meaningful aggregate quantities. Accordingly, the expected spatial association of an activity with all other activities gives a measure of the spatial `co-dependence' of an activity. This measure reveals how `picky' activities are in their tendencies to co-locate with other activities. This spatial co-dependence is low for activities that locate independently of other activities, whereas co-dependence is high for activities that are preferentially found in the presence of specific other activities. As an empirical illustration, we calculate the associations between pairs of occupations groups, along with the aggregate spatial co-dependence of each occupation group, using US city-occupation employment data. The associations between occupation groups reveal three clear clusters. The first consists of occupations related to knowledge intensive services, the second to occupations related to non-traded services and the third to occupations related to manufacturing. 



\section{Information-theoretic measures of (co-)location}\label{sec:intropmi}

\subsection{Notation}\label{sec:notation} 

Consider data on the location of economic activities in the form of an $N_c \times N_i$ dimensional matrix $\bQ$, where $N_c$ and $N_i$ are the number of locations and economic activities in the classifications of the data, respectively. We call $\bQ$ the `prevalence matrix' as its entries $q_{ci}$ denote the number of occurrences of activity $i$ in location $c$. This can be for example the number of people employed in a particular occupation $i$ in a city $c$, the number of establishments of industry $i$ in region $c$ or the number of dollars of product $i$ exported by country $c$. The total amount of activity of type $i$ and the total activity in location $c$ are given by the row sums $q_c = \sum_i q_{ci}$ and column sums $q_i = \sum_c q_{ci}$, respectively. Total economic activity is given by $q=\sum_{c,i} q_{ci}$. 

We will consider the prevalence matrix $\bQ$ to be the outcome of a sampling process from the underlying distribution $\bp$ with probabilities 
\begin{align} \label{eq:pci}
p_{ci}=P(X = i, C = c)
\end{align} 
that a randomly sampled unit (i.e. an employee, an establishment, a dollar) is part of activity $i$ in location $c$. Here, the categorical random variables $X$ and $C$ denote the activity and location of a randomly sampled unit, respectively. Their marginal probabilities are given by $p_i = \sum_c p_{ci} = P(X=i)$ and $p_c = \sum_i p_{ci} = P(C=c)$. 

The location-activity probabilities $p_{ci}$ will be the main object of interest as they hold information on the associations between locations and activities (Section \ref{sec:pmiloc}). From these probabilities it is also possible to construct the probabilities $p_{ij}$ that a pair of economic activities $i$ and $j$ are present in the same location, which is used to analyze the co-location association (Section \ref{sec:pmicoloc}). Both $p_{ci}$ and $p_{ij}$ are estimated from $\bQ$ using a Bayesian framework as described in Section \ref{sec:bayes}.

\subsection{Location association} \label{sec:pmiloc}
As noted, we will use the dependencies hidden in the joint probabilities $p_{ci}$ to measure the association between an activity and a location. Information theory provides a framework in which these associations can be quantified explicitly in units of information. The association between the two events $X=i$ and $C=c$ is given by their pointwise mutual information $PMI(p_{ci})$ \citep{Fano1961}. Intuitively, it answers the question `how much information does observing $c$ provide about the presence of $i$?' PMI has been used in several fields, including economics \citep{Theil1967}, administrative sciences \citep{Theil1972}, and linguistics \citep{Church1990}. Here, we use it in the context of economic geography to measure the association between economic activities and locations (location association) and within pairs of economic activities (co-location association). 

The PMI measures the association between two outcomes by assessing the information content of the realization $(C=c,X=i)$ given the information content in case of a null model in which $c$ and $i$ are independent, i.e. $p_{ci} = p_cp_i$. This is given by the logarithm of ratio of both probabilities:\footnote{In information theory, the information content or 'surprise' of an outcome $i$ is defined as $\log(\frac{1}{p_i})$. Observing an event that occurs with small probability leads to a high information content or surprise, whereas highly likely events contain little information. The difference between the information contents of $p_{ci}$ and $p_cp_i$ gives a measure of the surprise of observing $p_{ci}$ while expecting $p_cp_i$. Depending on the base of the logarithm, PMI measures association in units of bits (base 2) or nats (natural logarithm).}
\begin{align} \label{eq:PMI}
PMI(p_{ci}) = \log \left(\frac{p_{ci}}{p_cp_i}\right).
\end{align} 
$PMI(p_{ci})$ will be positive when it is more likely to observe $c$ and $i$ together than expected under independence, i.e. $p_{ci} > p_cp_i$, whereas $PMI(p_{ci})$ takes negative values when $c$ and $i$ are less likely to occur together than expected under the null model of independence, i.e. $p_{ci} < p_cp_i$. $PMI(p_{ci})=0$ if and only if $p_{ci} = p_cp_i$, indicating that $c$ and $i$ are independent (i.e., the incidence of an activity is independent of the place). The maximum value of $PMI(p_{ci})$ is given by $\max\{\log\left(\frac{1}{p_{i}}\right), \log\left(\frac{1}{p_{c}}\right)\}= \log\left(\frac{1}{p_{ci}}\right)$, which is attained either when activity $i$ always occurs in location $c$, or when activity $i$ is the only activity in location $c$.\footnote{Notice that then $p_{ci}=p_i$ or $p_{ci} = p_c$ respectively.} $PMI(p_{ci})$ is not bounded from below, as it tends to $-\infty$ as the joint probability $p_{ci}$ tends to $0$. 

\subsection{Co-location association} \label{sec:pmicoloc}
We can also use this information-theoretic framework to obtain a measure of association between pairs of economic activities. To do so, we expand \eqref{eq:pci} to include two units of activity:
\begin{align} \label{eq:pcij}
p_{cij}=P(X_1 = i, X_2 = j, C = c),
\end{align} 
where $X_1$ and $X_2$ describe randomly sampled units of activity from the same location $C$.

The measure of co-location will come from integrating across places to get the joint distribution of economic activities 
\begin{align} \label{eq:X1X2} 
p_{ij} = P(X_1 = i, X_2 = j).
\end{align} 
The probability $p_{ij}$ thus represents the joint probability that two units of economic activity that are randomly picked from the same (random) location are of type $i$ and $j$. It can be obtained by exploiting the fact that, conditional on knowing the location $c$, the occurrence of types $i$ and $j$ are independent, i.e. $p_{ij|c} = p_{i|c}p_{j|c}$, since the full distribution of economic activities for every location is known. By the law of total probability, one then obtains
\begin{align} \label{eq:pij}
p_{ij} = \sum_c p_{i|c}p_{j|c}p_c.
\end{align}
This defines the probability that two randomly sampled units from the same (random) location have activity types $i$ and $j$. 

As with the location-activity associations, the association between activity types can be quantified with the PMI. The association between two activities is then defined as
\begin{align}\label{eq:PMIpij} 
PMI(p_{ij}) = \log\left(\frac{p_{ij}}{p_ip_j}\right),
\end{align} 
where $p_ip_j$ is the null model that describes a situation where $i$ and $j$ are distributed independently of each other. What $PMI(p_{ij})$ captures is that the presence of some activities may increase or decrease the probability that other activities are present in the same location. Hence, observing a particular type of economic activity holds information about the likelihood of observing other types of activities in the same location. Economic activities that are more likely to occur together than expected under independence will have a positive association, whereas activities that are less likely to occur together than expected under independence will have a negative association.\footnote{Another way of seeing this, is by noting that $PMI(p_{ij})$ is positive when observing type $i$ increases the probability of observing type $j$ when sampling units of activity from the same location, i.e. $p_{j|i} > p_j$. Likewise, negative associations indicate that conditional on observing $i$, the probability of sampling a unit of activity $j$ in the same location decreases.} The $PMI(p_{ij})$ is inherently symmetric, since $p_{ij} = p_{ji}$. Computing this measure for all pairs of activity types thus leads to a symmetric, square matrix that has as entries the co-location association $PMI(p_{ij})$. 

The diagonal entries of this matrix hold `self-associations' $PMI(p_{ii})$. Self-association is high when observing an activity of type $i$ in a particular region increases the likelihood that a second randomly sampled unit in that location is also of type $i$. This is the case when the probability of observing $i$ is above average in a few locations, and below average in others. The self-association can thus be interpreted as a measure of geographical concentration. Note that the self-association is always positive, i.e. $PMI(p_{ii})\geq0$, since observing a unit of activity of type $i$ can never lower the probability of finding another unit of activity of type $i$ (we sample with replacement). The matrix of co-location associations thus provides a joint estimate of geographic concentration and co-location.



\section{Bayesian estimation} \label{sec:bayes}

In order to compute the quantities above, an estimate of the probabilities $p_{ci}$ is needed. A straightforward way to estimate these probabilities is to consider the share of every location-activity pair, corresponding to the maximum likelihood estimate $\p_{ci} = \frac{q_{ci}}{q}$. Here we estimate $p_{ci}$ using a Bayesian framework, which has two major advantages over the maximum likelihood approach. First, the Bayesian approach always returns nonzero probability estimates, so that computing the PMI will always return finite values. Second, the Bayesian framework yields a full posterior distribution for the estimated probabilities as opposed to a point estimate. The posterior distribution provides a natural description of the uncertainty in the estimated parameter values, which can be used to construct a Bayesian error bar for the information-theoretic quantities based on those estimates \citep{Wolpert1995}. 

Assuming that $\bQ$ is generated by an independent sampling process, the probability of its realization is given by a multinomial distribution
\begin{align*} 
P(\bQ | \bp) = \frac{\Gamma(q+1)}{\prod_{c,i} \Gamma(q_{ci}+1)} \prod_{c,i} p_{ci}^{q_{ci}},
\end{align*}
where $\bp$ is the matrix containing probabilities $p_{ci}$, $\sum_{c,i}p_{ci} = 1$.

Applying Bayes' rule, the posterior distribution for the matrix of probabilities $\bp$ is then given by
\begin{align*} 
P(\bp | \bQ) \propto P(\bQ | \bp)P(\bp),
\end{align*}
where $P(\bp)$ represents the prior distribution. A conjugate prior for the multinomial distribution is the Dirichlet distribution 
\begin{align*} 
P(\bp | \bm{\alpha}) \sim Dir(\bm{\alpha}) = \frac{\Gamma(\alpha)}{\prod_{c,i} \Gamma(\alpha_{ci})}\prod_{c,i} p_{ci}^{\alpha_{ci}-1},
\end{align*} 
where $\alpha = \sum_{c,i}\alpha_{ci}$. This gives the distribution of $\mathbf{p}$ given hyperparameter $\bm{\alpha}$. The posterior distribution for $\bp$ given the data $\bQ$ and hyperparameter $\bm{\alpha}$ is then given by
\begin{align*}  
P(\bp | \bQ,\bm{\alpha}) \sim Dir(\bQ + \bm{\alpha}) \propto \prod_{ci} p_{ci}^{q_ci + \alpha_{ci}-1} .
\end{align*} 

The hyperparameter $\bm{\alpha}$ can be interpreted as a matrix of `pseudocounts', giving the assumed number of observed units of activity for every $c,i$ pair prior to seeing the data $\bQ$. The total number of pseudocounts $\alpha$ determines the strength of the prior relative to the data. 
An estimate for the parameters $p_{ci}$ is then given by the expectation of the marginals of the posterior distribution, so that 
\begin{align*}
\p_{ci} = \E[p_{ci}|\bQ, \bm{\alpha}] = \frac{q_{ci}+\alpha_{ci}}{q + \alpha} = \frac{\q_{ci}}{\q},
\end{align*}
where we write $\q_{ci} = q_{ci} + \alpha_{ci}$ and $\q = \alpha + q$. When the pseudocounts $\alpha_{ci}$ are nonzero for all $c,i$, then $\p_{ci}>0$ will also be nonzero. This has the practical advantage that it prevents difficulties when computing logarithms of the estimated probabilities, as when calculating $PMI(p_{ci})$.\footnote{In the context of information retrieval in text analysis, adding the pseudocounts $\alpha_{ci}$ to categorical data is known as `Laplace smoothing' or `additive smoothing' \citep{Manning2008}.}

A measure for the uncertainty of the estimate $\p_{ci}$ is given by the variance of the marginals of the posterior distribution, leading to 
\begin{align*} 
\Var[p_{ci}|\bQ, \bm{\alpha}] &= \frac{\q_{ci}(\q - \q_{ci})}{\q^2(\q+1)}\\
&=\frac{\q_{ci}/\q(1 - \q_{ci}/\q)}{\q+1}.
\end{align*} 
Note that this implies that the variance is dependent on the granularity of the data in $\bQ$. To see this, suppose we alter the units leading to a new matrix $\bQ' = k \bQ$, so that for large $q$
\begin{align*}
\Var[p_{ci}|\bQ', \bm{\alpha}] 
&= \frac{k\q_{ci}/k\q(1 - k\q_{ci}/k\q)}{k\q+1}\\
&\approx \frac{1}{k} \Var[p_{ci}|\bQ, \bm{\alpha}].
\end{align*}
The variance thus decreases as the counts become more fine-grained. The reason is that the data generating process is assumed to create the data at the level of the counts, so that more-fine grained units represent more observations. The variance of the estimates is thus directly related to the units in which the underlying data generating process is assumed to generate the data.\footnote{In the context of (co-)agglomeration of industries for example, the relevant unit of analysis is the one at which location decisions are made, which could be be assumed to be the plant level, suggesting an analysis of data containing the counts of plants of a specific industry for a given location.} However, the variance is affected by the granularity of the data in the same way across activities and locations, so that the \emph{relative} uncertainty of estimates $\p_{ci}$ is independent of the units of $\bQ$. 

One could use the estimate $\p_{ci}=\E[p_{ci}]$ directly to compute $PMI(\p_{ci})$ and $PMI(\p_{ij})$. However, this will induce a systematic bias which comes from Jensen's inequality $\E[PMI(p_{ci})] \lesseqgtr PMI(\E[p_{ci}])$ depending on whether $PMI(p_{ci})$ is concave or convex.\footnote{$PMI(p_{ci})$ is concave when $\partial^2 PMI(p_{ci})/\partial p_{ci}^2 = -1/p_{ci}^2 + 1/p_c^2 + 1/p_i^2 < 0$, and convex when $\partial^2 PMI(p_{ci})/\partial p_{ci}^2 > 0$.} One needs instead an estimate of $PMI(p_{ci})$, which in itself is a random variable whose distribution is determined by the posterior distribution of $p_{ci}$. Thus, we use the uncertainty for the estimates $\p_{ci}$ to determine the uncertainty of estimates for $PMI(p_{ci})$ and $PMI(p_{ij})$.

\subsection{Estimation of the posterior mean and variance of $PMI(p_{c,i})$}

Here, we approximate the mean and variance of the posterior distribution of $PMI(p_{ci})$, which will serve as estimates of the posterior distribution of the location-activity association. Our approach is based on \cite{Wolpert1995} and \cite{Hutter2005}, in which the estimation of information-theoretic quantities using a Bayesian approach is discussed in depth. 

To obtain an approximation for the posterior distribution of $PMI(p_{ci})$, we compute its Taylor expansion around the mean $\p_{ci}$. Writing $\Delta_{ci} = p_{ci} - \p_{ci}$, and noting the fact that $\left|\Delta_{ci}\right| < 1$, this gives
\begin{align*}
PMI(p_{ci}) &= PMI(\p_{ci}) + \Delta_{ci}\left(\frac{1}{\p_{ci}} - \frac{1}{\p_c} -\frac{1}{\p_i}\right)    \nonumber   \\
&+ \frac{\Delta_{ci}^2}{2}\left(-\frac{1}{\p_{ci}^2} + \frac{1}{\p_c^2} + \frac{1}{\p_i^2}\right) + \mathcal{O}(\Delta_{ci}^3). 
\end{align*}
Note that $\E[\Delta_{ci}] = 0$ and thus $\E[\Delta_{ci}^2] = \Var[p_{ci}]$, where expectations are taken with respect to the posterior distribution of $p_{ci}$. It follows that
\begin{align} \label{eq:EPMIci}
\E[PMI(p_{ci})] \approx PMI(\p_{ci}) + \frac{\Var[p_{ci}]}{2}\left(-\frac{1}{\p_{ci}^2} + \frac{1}{\p_c^2} + \frac{1}{\p_i^2}\right).
\end{align}
The second term accounts for systematic bias in the estimate of $PMI(p_{ci})$, in which the sign of the factor multiplying the variance is indicative of whether $PMI(p_{ci})$ is concave or convex, and thus determines whether the bias is positive or negative. 

Using the Delta method, we then obtain for the variance of $PMI(p_{ci})$:
\begin{align} \label{eq:varPMIci}
\Var[PMI(p_{ci})] &\approx \Var[p_{ci}]\frac{\partial PMI(p_{ci})}{\partial p_{ci}}|_{\hat{p}_{ci}} \nonumber \\
&= \Var[p_{ci}]\left(\frac{1}{\p_{ci}} - \frac{1}{\p_c} - \frac{1}{\p_i} \right)^2.
\end{align}
This is a measure for the uncertainty around the point estimate $\E[PMI(p_{ci})]$. In particular, it can be used to determine whether the estimate for $PMI(p_{ci})$ is significantly nonzero, i.e. if there is a significant association between $i$ and $c$.

\subsection{Estimation of posterior mean and variance of $PMI(p_{ij})$} 

Approximations of $\E[PMI(p_{ij})]$ and $\Var[p_{ij}]$ are obtained in a similar fashion, replacing $p_{ci}$ with $p_{ij}$ in equations \eqref{eq:EPMIci} and \eqref{eq:varPMIci}, although the computation of $\Var[p_{ij}]$ is more involved. \ref{app:varpij} provides a discussion of how $\Var[p_{ij}]$ is obtained. \ref{app:MCMC} provides comparisons to numerical simulations to justify the approximations made. 

\section{Location and co-location} \label{sec:location} 

\subsection{Revealed Comparative Advantage}

One of the most commonly used indices to study location patterns of economic activities originates from the trade literature, where it is known as Balassa's index of Revealed Comparative Advantage (RCA) \citep{Balassa1965}. The $RCA$ of a location-activity pair is given by the ratio of the share of activity $i$ within location $c$ compared to the share of activity $i$ in the overall economy:
\begin{align} \label{eq:RCA}
RCA(c,i) = \frac{q_{ci}}{q_c} / \frac{q_i}{q}.
\end{align} 
It compares the observed share of activity $i$ within location $c$ in the numerator to the total share of $i$ as given by the denominator. Since $q_i$ and $q_c$ are exchangeable in \eqref{eq:RCA}, $RCA(c,i)$ can be interpreted in two ways: as a measure of `localization' of activity $i$ in location $c$, or as a measure of `specialization' of location $c$ in activity $i$. The neutral value is given by $RCA(c,i)=1$, where the share of activity $i$ in location $c$ is equal to the total share of activity $i$ over all locations. 

A theoretical derivation of the $RCA$ index is given by \cite{Kunimoto1977}, who uses a probabilistic approach that comes close to the approach presented in this paper. Properties of the $RCA$ and related indices have since been discussed extensively \citep{Yeats1985,Ballance1987,Vollrath1991}, some of which are problematic when applying the index in empirical analysis. One of the issues of the $RCA$ index is that it is heavily skewed and asymmetric around its neutral value. A possible solution that has been presented is taking the logarithm of the index, making it symmetric around a neutral value of $0$ \citep{Vollrath1991}. This, however, leads to the problem that the index becomes undefined in the cases where $q_{ci}=0$, since the logarithm of zero is undefined. 



The approach presented in the current paper provides an information-theoretic derivation of the logarithm of the RCA index. Consider the maximum likelihood estimate for the multinomial probabilities $\p_{ci}=\frac{q_{ci}}{q}$.\footnote{Note that here we write $q_{ci}$ and not $\q_{ci}= q_{ci} + \alpha_{ci}$, since the maximum likelihood estimate uses directly the observed counts, without adding the pseudocounts that where a consequence of incorporating a prior distribution of counts in the Bayesian estimate.} We then have that 
\begin{align*} 
PMI(p_{ci}) &= \log\left(\frac{\p_{ci}}{\p_c \p_i}\right) \\
&= \log\left(\frac{\q_{ci}}{\q_c}\bigg/\frac{\q_i}{\q}\right) \\
&= \log(RCA(c,i)),
\end{align*}
showing that conceptually the PMI is equal to the logarithm of the $RCA$ index. 

Our approach stands therefore as a generalization of the $RCA$ index. This shows that there is an information-theoretic notion of association underlying the $RCA$. Seen in this light, the practical problem of having to take the logarithm of zero when $q_{ci}=0$ is in fact a problem related to miss-estimating $p_{ci}$. In our Bayesian approach, the estimates of probabilities $p_{ci}$ are always strictly positive. 

\subsection{Measures of localization} \label{sec:localization}


Many questions are better answered at more aggregate levels of analysis than the level of location-activity pairs. Typical questions at these levels of aggregations rely on quantifying which activities are most localized in space, or which locations are most specialized in terms of their economic activities. 

Localization of an activity can be defined as the degree of dissimilarity between the activity's own geographical distribution and the distribution of the population or of total economic activity across all locations \citep{Hoover1936, Mori2005}. Highly localized activities will be distributed across locations in a very different way than what one would expect from locations' sizes. Activities with a low degree of localization will be distributed proportionally to the relative (population) size of locations. This can be quantified by comparing how much, on average, the probability that a unit of activity of type $i$ is located in a location differs from the probability that any unit of activity is located there. 

Let $p_{c|i}=p_{ci}/p_i$ be the probability that a unit of activity is located in $c$ given that its activity type is $i$, and recall that the probability that a unit of economic activity is located in $c$ regardless of its type is given by $p_c$. Considering the average deviations between $p_{c|i}$ and $p_c$ leads to a measure of localization that is given by
\begin{align*} 
KL(p_{c|i} | p_c) &= \sum_c p_{c|i}\log(p_{c|i}/p_c)\\
&= \sum_c p_{c|i} PMI(p_{ci}),
\end{align*} 
where we used that $p_{c|i}/p_c = p_{ci}/p_cp_i$. Here, $KL$ denotes the Kullback-Leibler divergence \citep{Kullback1951}, and measures the deviation between the distribution across all locations of a specific activity, given by probabilities $p_{c|i}$, and the overall distribution of locations, given by the probabilities $p_c$. Hence, the proposed information-theoretic framework naturally suggests a localization measure by aggregating $PMI(p_{ci})$ to the activity level. The resulting metric can be interpreted as the activity type's expected locational dependence. 


This measure has the exact same functional form as the measure of industrial localization put forward by \cite{Mori2005}, although the null model implicit in their metric is based on a location's area. That is, they take the probability $p_c$ to be proportional to the area of that location as opposed to its population size.\footnote{Furthermore, they obtain an error bar for this statistic based on a normal approximation. In the Bayesian framework, an estimate for the standard deviation of the $KL$ can be obtained in a similar way as for the PMI, as shown in \ref{app:varKL}.} Here we show that their measure can be retrieved as the expected PMI values of a particular industry.\footnote{This holds regardless of the 'null model' considered. Hence, one could follow \cite{Mori2005} and use their area based null model to define a measure on the location-activity level that is analogous to the $RCA$ index.} Ignoring differences in how these distributions are estimated, the functional of this measure is equal to $\E_{p_{c|i}}[\log(RCA(c,i))]$, showing that it can be understood as the expected value of the logarithm of the $RCA$ of an activity over all locations it occurs in. 

\subsection{Measures of specialization} 

Similarly, the aggregate level of specialization of a location as a whole can be analyzed by quantifying the difference of the distribution of activities within the location, $p_{i|c}$, to the overall distribution of activities $p_i$. Such a measure of specialization is obtained by aggregating the $PMI(p_{ci})$ to the location level, thus considering the expected association of the activity with particular locations, leading to
\begin{align*} 
KL(p_{i|c} | p_i) = \sum_c p_{i|c} PMI(p_{ci}).
\end{align*}
Again, this can be interpreted as the expected value of the logarithm of the $RCA$, but now over industries within a given location: $\E_{p_{i|c}}[\log(RCA(c,i))]$. The measure is akin to Krugman's specialization index \citep{Krugman1991b}.\footnote{The Krugman specialization index is given by $K(c) = \sum_i |p_{i|c} - p_i|$. Like $KL(p_{c|i}|p_i)$, it considers an 'average deviation' of $p_{i|c}$ to $p_i$, where the measure of deviation is taken to be the absolute difference.} However, in our framework, the localization of activities and specialization of locations are essentially the same measures, defined for different units of analysis. 

\subsection{Overall specialization}

Aggregating even further, a measure for the overall specialization at the system level can be obtained by taking the expectation over both locations and activities, leading to the expected association of a location-activity pair, or equivalently as either the expected localization of an activity or the expected specialization of a location. The resulting quantity is known as the Mutual Information (MI) \citep{Cover2005} and quantifies the dependence between two random variables. In this case, it measures the dependence between the random variables $X$ and $C$, which describe the type and location of a randomly sampled unit of activity. It is given by
\begin{align} \label{eq:MI}
MI(C,X) &= \sum_{c,i} p_{ci} PMI(p_{ci})\\
&= \sum_i p_i KL(p_{c|i}|p_c)\\
&= \sum_c p_c KL(p_{i|c}|p_i).
\end{align} 

When $MI(C,X)=0$, the location of a randomly sampled unit is independent of its activity type, which implies that all economic activity is distributed proportionally to location size, or equivalently that every location has an identical distribution of activities. In this situation, there is no specialization in the system in the sense that all locations are identical. The maximum value of $MI(C,X)$ is reached when each location has its own unique activity, so that each location is maximally specialized and each activity is maximally localized. In the current context, the mutual information is a system-level measure of overall specialization that can be used to compare across different systems (e.g. comparing the degree of overall specialization across countries), or to track the changes over time (e.g. comparing the degree of overall specialization before and after the establishment of a trade union). Table \ref{tab:actloc} summarizes each of the measures derived thus far and the relation between them. 


\begin{table}
  \resizebox{\columnwidth}{!}{
 \begin{tabular}{|l|l|c|}
 \hline
      unit of analysis & measure & formula \\
      \hline 
      location-activity & association & $PMI(p_{ci})$ \\
      activity & localization & $KL(p_{c|i}|p_c) = \E_{p{c|i}}[PMI(p_{ci})]$ \\
      location & specialization & $KL(p_{i|c}|p_i) = \E_{p{i|c}}[PMI(p_{ci})]$\\
      system & overall specialization & $MI(C,X) = \E_{p{ci}}[PMI(p_{ci})]$\\
    \hline 
\end{tabular}}
\caption{}
\label{tab:actloc}
\end{table}

\section{Co-location}

\subsection{Co-location association} 
So far, we have studied the matrix $\bQ$, which summarizes location patterns of economic activity. Our framework can however readily be extended to study more complex patterns. Here we will discuss co-location patterns of pairs of activities, i.e., of the dependencies between activities that are located in the same region. Such co-location patterns have received increasing attention in studies on international trade \citep{Hidalgo2007} and urban economies \citep{Ellison2010}. In the latter field, authors have used co-location patterns to test theories on Marshallian externalities \citep{Marshall1920}. In this literature, the co-agglomeration index of \citet{Ellison2010} has become a de facto standard \citep{Faggio2017, Diodato2018}. Here, we show how information theory can be used to derive an alternative measure based on the co-location association, $PMI(p_{ij})$.

Before presenting our co-location metrics in detail, it is useful to first discuss how \citet{Ellison2010} construct their co-agglomeration index. These authors present a location choice model for profit-maximizing plants \citep{Ellison1997, Ellison2010} in which the (combined) effects of natural advantage and spillovers between activity types determine co-agglomeration patterns. They propose the following pairwise co-agglomeration index:\footnote{Note that, in our notation, activity shares $\frac{q_{ci}}{q_i}$ and $\frac{q_c}{q}$ are replaced by probabilities $p_{c|i}$ and $p_c$. This makes specific that we regard the former shares as maximum likelihood estimates of the latter probabilities. For now, however, we leave the issue of estimating these probabilities open.}
\begin{align}\label{eq:GAMMAij}
    \gamma_{ij}  =  \frac{\sum_c (p_{c|i} - p_c)(p_{c|j} - p_c)}{1-\sum_c p_c^2}. 
\end{align}

The co-agglomeration of all pairs can be collected in a matrix with entries $\gamma_{ij}$, completely analogous to the $PMI(p_{ij})$ in Section \ref{sec:pmicoloc}. The diagonal entries $\gamma_{ii}$ contain the agglomeration index of a single activity \citep{Ellison1997}, when neglecting effects of the plant size distribution.\footnote{\cite{Mori2005} show that the agglomeration index of \citep{Ellison1997} can be written as $\gamma_i = a_i G_i - b_i \approx \frac{\sum_c (p_{c|i} - p_c)^2}{1-\sum_c p_c^2}$. This approximation is valid when plants are reasonably uniformly distributed, in which case the plant size effect is negligible. The plant size distribution determines the size of the chunks in which the counts are generated in the data generating process. Quantifying the dependencies that arise from such a data generating process is an interesting direction for future research, but for now we focus on the simpler case in which information on the chunk sizes (e.g. the plant size distribution) is unavailable. Further note that unlike \cite{Mori2005}, we compare the agglomeration index to the self-association $PMI(p_{ii})$ as opposed to the localization $KL(p_{c|i}|p_c)$.} 

Comparing the co-agglomeration index given in Eq. (\ref{eq:GAMMAij}) to our co-location association metric rewritten as
\begin{align*}
    PMI(p_{ij}) &= \log\left(\frac{\sum_c p_{i|c}p_{j|c} p_c}{p_ip_j} \right) \\
    &= \log\left( \sum_c \left(\frac{p_{c|i}}{p_{c}}\right)\left(\frac{p_{c|j}}{p_{c}}\right)p_c \right).
\end{align*}
clarifies the conceptual similarity between the two. Both capture how different activities co-vary in space. In either case, the intensity of spatial co-location may be generated by a location choice model akin to the one by \cite{Ellison1997}. 

The difference lies, however, in the functional form used to measure the deviation from the reference distribution. The co-location association compares probabilities by taking ratios $p_{i|c}/p_c$, whereas the co-agglomeration index considers differences $p_{i|c} - p_c$. Furthermore, the co-location association weights each of the differences by $p_c$. 

Although the co-agglomeration index is derived from an economic model, the measure of concentration that lies at its heart enters the derivation as an assumption. Our framework provides a principled way to quantify these deviations, by leveraging information theory. The advantage of such an approach is that it gives insight into the underlying assumptions on the data generating process, the used reference distribution\footnote{In fact, the literature is not entirely consistent in the choice of the reference distribution that is used in the (co-)agglomeration indices. In some work the reference distribution is taken to be the share of total employment in location $c$, which we denote by $p_c$ \citep{Ellison1997, Ellison1999, Faggio2017}. In other work, the reference distribution is given by the average share of employment in industry $i$ in a location, given by $\hat{p}_{c|i} = \frac{1}{N_i}\sum_i p_{c|i}$ \citep{Ellison2010, Diodato2018}.}, and the estimation procedure with its corresponding uncertainties. Furthermore, as before, our statistical framework allows constructing measures of co-dependence at higher levels of aggregation, such as at the level of the activity or of the economic system as a whole. 



\subsection{Co-dependence} 
As in Section \ref{sec:localization}, the co-location associations can be aggregated by taking the expectation across all activities $j$, leading to a measure of the average association of activity $i$ with all other activities, given by 
\begin{align}\label{eq:KLpij}
    KL(p_{j|i}|p_j) = \sum_j p_{j|i}PMI(p_{ij}).
\end{align}
We call this measure the co-dependence of a particular activity. It quantifies the deviation of the distribution of activity types conditional on having observed activity type $i$, $p_{j|i}$, with respect to the unconditional distribution of probabilities $p_j$. When activity type $i$ has, on average, strong associations with other activity types it co-locates with, this deviation will be large. In other words, activity $i$ `cares' about the type of activity it co-locates with. A low value of $KL(p_{j|i}|p_j)$ on the other hand implies that the distribution of probabilities $p_{j|i}$ does not differ much from the distribution of $p_j$, meaning that activity $i$ is uninformative for the type of activities it co-locates with. This implies that activity $i$ co-locates with the 'average' distribution of activity types, suggesting it is indifferent of the other activities in the same location. 

Note that activities that are heavily concentrated geographically, have by definition a high co-dependence, as $PMI(p_{ii})$ is part of the sum in \eqref{eq:KLpij}. In that case, activity of type $i$ typically co-locates with other activity of type $i$.

\subsection{Overall pairwise dependence}
Taking the expectation of the co-dependence over all activity types, or equivalently taking the expectation of the co-location association over all activity pairs leads to the mutual information
\begin{align*} 
MI(X_1,X_2) &= \sum_i p_i KL(p_{j|i}|p_j)\\
&= \sum_{ij} p_{ij} PMI(p_{ij}).
\end{align*} 

This is a measure of dependence between the random variables $X_1$ and $X_2$, which each describe the activity type of a randomly sampled unit of activity, both sampled from the same location (see \eqref{eq:X1X2}). The overall co-dependence is thus a system-level variable that describes how much two units of activity are on average (spatially) associated. This may, for instance, help understand how the overall strength of co-agglomeration externalities differs across economies or changes over time. Table \ref{tab:actact} gives a summary of the measures that follow from analysis of the co-location distribution $p_{ij}$. Both Tables \ref{tab:actloc} and \ref{tab:actact} construct similar sets of measures. Both sets of measures take averages across rows, columns, or both, of a matrix that summarizes associations between two variables. However, whereas the measures in Table \ref{tab:actloc} are based on the location-activity information of a matrix that collects elements $PMI(p_{ci})$, the measures in Table \ref{tab:actact} are based on the spatial co-location information collected in a matrix with elements $PMI(p_{ij})$. 

\begin{table}
 \resizebox{\columnwidth}{!}{
 \begin{tabular}{|l|l|c|}
 \hline 
      unit of analysis & measure & formula \\
      \hline 
     activity-activity & co-location association & $PMI(p_{ij})$\\
     activity-activity & geographic concentration & $PMI(p_{ii})$\\
     activity & co-dependence & $KL(p_{j|i}|p_j) = \E_{p_{j|i}}[ PMI(p_{ij})]$\\
     system & overall co-dependence & $MI(X_1,X_2) =  \E_{p_{ij}}[PMI(p_{ij})]$\\
     \hline 
\end{tabular}}
\caption{}
\label{tab:actact}
\end{table}


\section{Empirical example}

As an example, we apply the PMI to show the co-location associations of occupation groups in US employment data in 2016 provided by the Bureau of Labor Statistics.\footnote{These data are available at \url{https://www.bls.gov/oes/special.requests/oesm16ma.zip}} The data consists of a matrix $\bQ$ that gives for every city $c$ the number of employees $q_{ci}$ in a particular occupation group $i$. In this example, we choose a uniform prior, setting $\alpha_{ci}=1$ for all $c,i$. This represents a single observation for every location-activity pair. Since the total number of pseudocounts $\alpha = N_rN_c << q$, the resulting estimates will be determined much more by the data than by than the prior. 

\begin{figure*}[h]
\centering
\includegraphics[width=\linewidth]{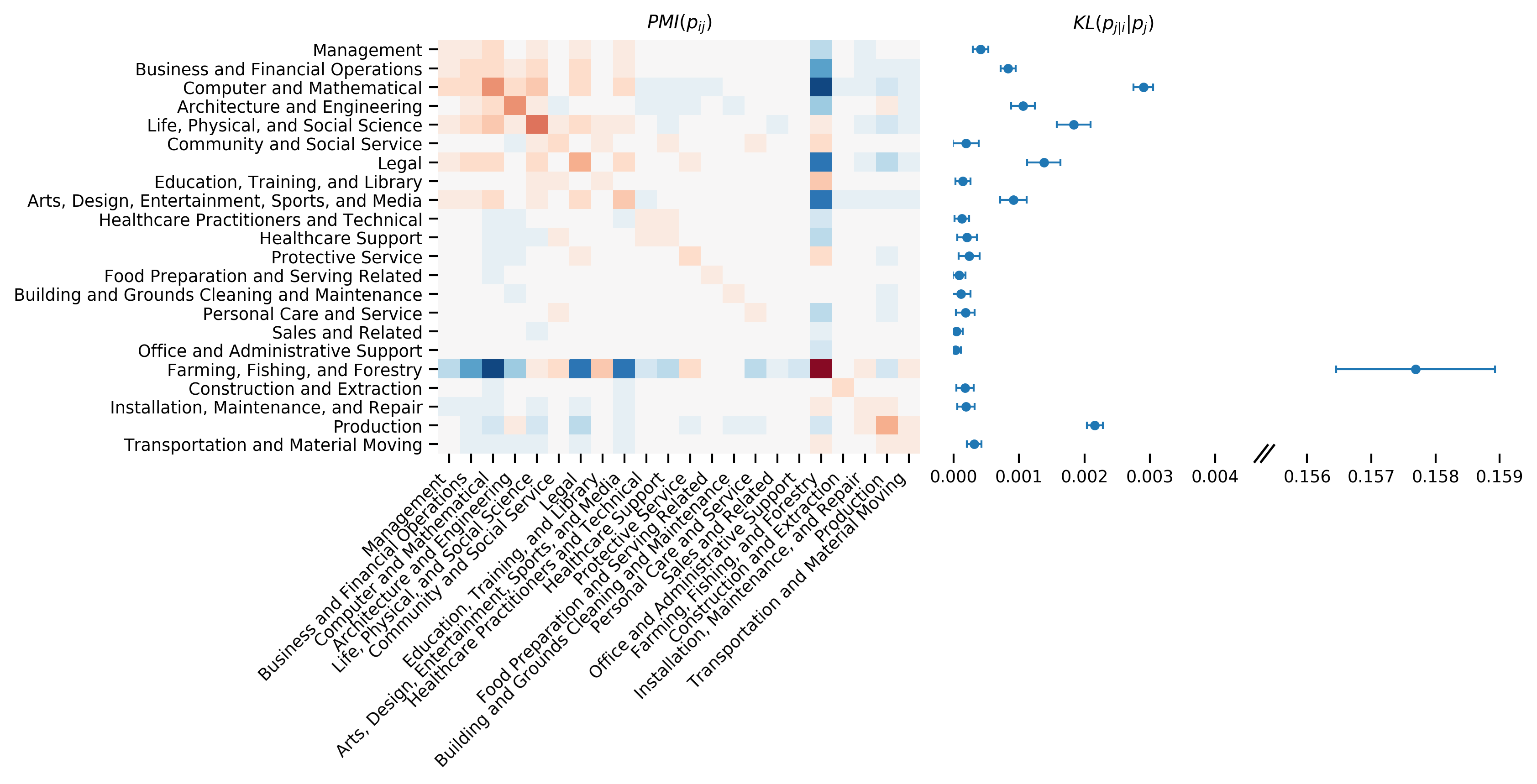}
 \caption{Values of the estimated $PMI(p_{ij})$ for major occupations groups. Red indicates positive associations, blue indicates negative associations, and grey indicates neutral ($PMI(ij)=0$) associations. All pairwise associations are between $-0.5$ and $0.5$ with the exception of the self-association of the 'Farming, fishing and forestry' occupations, which has a value of $3.15$. The right hand side shows the co-dependence $KL(p_{j|i}|p_j)$ of every occupation group, given by the expected value of a row of the $PMI$ matrix. The error bars depict one standard deviation of the posterior distribution as a measure of uncertainty for the estimate. Note the broken axis, showing the extreme dependence of the 'Farming, fishing and forestry' occupations.} 
\label{fig:PMIij}
\end{figure*}

The inferred $PMI(i,j)$ matrix is shown in Figure \ref{fig:PMIij}, showing the co-location associations between the occupation groups. The right hand side shows the co-dependence of each occupation group with respect to all other groups, corresponding to the expected value of a row in the PMI matrix. The error bars show one standard deviation in the posterior distribution, as derived in \ref{app:varKL}. Red indicates positive associations, and blue negative ones. 

The matrix delineates three clusters of occupations groups. The upper left block shows a cluster of positively associated occupations that seem to be related to knowledge-intensive services. The positive associations lead to a relatively high co-dependence for these occupations, suggesting that the presence of these occupations depends largely on which other occupations are present in the same city. 

The lower right block of the matrix shows a smaller cluster of occupations related to production, transportation and repair. These occupations have a negative association with the knowledge-intensive occupations, and thus typically co-locate with a different set of occupations. The `Production' occupations group also has a high co-dependence, which is mostly driven by a high self-association.

The 'Farming, fishing and forestry' group is highly isolated, with mostly negative associations with other groups. The diagonal entry in the matrix shows the self-association is very high, which is also reflected in a high co-dependence, which is orders of magnitude larger than that of the other occupations (note the broken axis). 

In the middle band of the matrix, occupation groups have a neutral association with most other occupations, and have a low co-dependence. These groups seem to be related to non-traded services, including 'Protective service', 'Food preparation and serving' and 'Personal care and service'. The low co-dependence implies that these occupations are distributed approximately proportional to the total population, independently of which other occupation groups are present in a city.



\section{Discussion}

Information theory offers a unified way to estimate location and co-location associations using PMI. This yields measures that are similar to the well-known RCA index \cite{Balassa1965} and the co-agglomeration index \citep{Ellison2010}. However, our metrics based in information theory have important advantages over these existing measures. 

First, by deriving these metrics from a unified framework, we were able to show the intrinsic connections between hitherto disparate measures. This is not only satisfying from a methodological point of view, but allows exploring the relations between concepts like revealed comparative advantage, specialization, localization, concentration and co-location. 

Second, the proposed measures are derived from a formal framework (information theory) in a way that is explicit in the assumed data generating process, the chosen null models and the estimation procedures. Different choices for these assumptions leads to different results. However, the afforded transparency allows to construct arguments against and in favor of such alternatives that take into consideration aspects of the specific context at hand. Such a discussion can be framed in terms of an underlying model, rather than of ad hoc specificities of a particular index. For instance, we used a null model based on the assumption that neutral associations imply a distribution of location-activity pairs that is proportional to the sizes of locations and activities \citep{Hoover1936}. Alternative null models could follow from the assumption that activities are distributed proportional to the area of a location \citep{Mori2005}. Another possibility is to determine the expected number of (co-)occurrences on the basis of external factors that could drive the distribution of activities over locations, using for instance a regression model \citep{Neffke2011, Jara-Figueroa2018}. 

Third, the framework provides uncertainty estimates for all the information-theoretic quantities involved. Most currently used indices are applied without any notion of uncertainty. Using these uncertainties in practice however may present some challenges. For instance, the Bayesian estimation procedure leaves room for the selection of different priors. Here, for reasons of practicality, we applied a simple uniform prior. However, in some contexts, alternative priors may be natural choices. Many of these priors would still result in Dirichlet priors, but with different uniform values for $\alpha_{ci}$ to adapt the strength of the prior to the data at hand \citep{Hutter2005}. In other contexts, non-uniform priors, such as the maximum entropy prior \citep{Wolpert1995}, may be preferable. Furthermore, the absolute magnitude of the uncertainty will depend on the granularity of the data. This simply reiterates that inferences should always be made with an underlying data generating process in mind. In spite of this, we can still make statements about the relative magnitudes of uncertainties, which are independent of the granularity of the data generating process.


Fourth and finally, it is important to note that the information-theoretic approach can be readily extended to move beyond the analysis of pairwise co-locations, as it also allows analyzing multivariate associations. For instance, one could analyze associations between multiple variables (e.g. occupations, cities and industries) or multi-way co-locations (such as the co-location of triplets instead of pairs of activities).\footnote{The PMI between three economic activities $i,j,k$ is given by $PMI(p_{ijk}) = \log\left(\frac{p_{ijk}}{p_ip_jp_k}\right)$.} Such higher-order associations could be further analyzed using the information-theoretic concepts of redundancy and synergy \citep{Finn2018}. This may help disentangle different types of associations, capturing different economic interactions. The association between a pair of economic activities could be conditional on the presence of (a specific combination of) other activities, or be driven by the mutual dependence on a (combination of) other economic activities or on some external variable such as the presence of a natural resource. Further development of this analytical framework could reveal such higher-order relations among economic activities.

\section*{Acknowledgements}
A.D and K.F are funded by the Netherlands Organisation for Scientific Research (NWO) under the Vici scheme, number 453-14-014.

\bibliographystyle{model2-names}
\bibliography{library}

\clearpage
\appendix
\onecolumn

\section{Estimation of $\Var[p_{ij}]$}\label{app:varpij}

Estimates for the expectation and variance of $PMI(p_{ij})$ are obtained in a similar fashion as \eqref{eq:EPMIci} and \eqref{eq:varPMIci}. This requires computation of $\Var[p_{ij}]$. We have 
\begin{align}\label{eq:varpij} 
\Var[p_{ij}] &= \Var[\sum_c p_{i|c}p_{j|c} p_c] \nonumber \\
&= \sum_c \Var[p_{i|c}p_{j|c} p_c] + \sum_{c \neq c'} \Cov[p_{i|c}p_{j|c}p_c,p_{i|c'}p_{j|c'}p_{c'}] \nonumber \\
&= \sum_c  \left(\Var[p_{i|c}p_{j|c}]\E[p_c]^2 + \Var[p_c]\E[p_{i|c}p_{j|c}]^2 + \Var[p_{i|c}p_{j|c}]\Var[p_c]\right) + \sum_{c \neq c'} \E[p_{i|c}p_{j|c}]\E[p_{i|c'}p_{j|c'}] \Cov[p_c,p_{c'}],
\end{align} 
where in the second equality we used that $p_{i|c}p_{j|c}$ is independent of $p_c$, $p_c'$ and $p_{i|c'}p_{j|c'}$. Furthermore, we used that 
\begin{align*}
  \Cov[p_{i|c}p_{j|c}p_c,p_{i|c'}p_{j|c'}p_{c'}] &= \E[p_{i|c}p_{j|c}p_cp_{i|c'}p_{j|c'}p_{c'}] - \E[p_{i|c}p_{j|c}p_c]\E[p_{i|c'}p_{j|c'}p_{c'}] \\
  &= \E[p_{i|c}p_{j|c}]E[p_{i|c'}p_{j|c'}](\E[p_cp_{c'}] - \E[p_c]\E[p_{c'}])\\
  &= \E[p_{i|c}p_{j|c}] \E[p_{i|c'}p_{j|c'}] \Cov[p_c,p_{c'}].
\end{align*}
Note that the vector of $p_c$'s follows a Dirichlet distribution, so that $p_c$ and $p_{c'}$ are not independent.

Using the product-moment formula \citep{Nadarajah2004}, we know that for $i \neq j$ 
\begin{align*}
\E[p_{i|c}^np_{j|c}^n] = \frac{\Gamma(\q_{ci}+n)\Gamma(\q_{cj}+n)\Gamma(\q_c)}{\Gamma(\q_{ci})\Gamma(\q_{cj})\Gamma(\q_c+2n)},
\end{align*}
so that
\begin{align*} 
\Var[p_{i|c}p_{j|c}] &= \E[p_{i|c}^2p_{j|c}^2] - \E[p_{i|c}p_{j|c}]^2\\
&= \frac{\q_{ci}(\q_{ci}+1)\q_{cj}(\q_{cj}+1)}{\q_c(\q_c+1)(\q_c+2)(\q_c+3)} -  \left(\frac{\q_{ci}\q_{cj}}{\q_c(\q_c+1)}\right)^2.
\end{align*} 
The last term of \eqref{eq:varpij} consists of 
\begin{align*} 
\sum_{c \neq c'} \E[p_{i|c}p_{j|c}]\E[p_{i|c'}p_{j|c'}] \Cov[p_c,p_{c'}] = - \sum_{c\neq c'} \frac{\q_{ci}\q_{cj}}{\q_c(\q_c+1)}\frac{\q_{c'i}\q_{c'j}}{\q_c(\q_c+1)}\frac{\q_c\q_{c'}}{(\q+1)}.
\end{align*}

For $i =j$ we have 
\begin{align*} 
\Var[p_{ii}] = \sum_c  \left(\Var[p_{i|c}^2]\E[p_c]^2 + \Var[p_c]\E[p_{i|c}^2]^2 + \Var[p_{i|c}^2]\Var[p_c]\right) + \sum_{c \neq c'} \E[p_{i|c}^2]\E[p_{i|c'}^2] \Cov[p_c,p_{c'}].
\end{align*} 
Since $p_{i|c}$ is beta-distributed, we have 
\begin{align*} 
\E[p_{i|c}^2] = \frac{\q_{ci}(\q_{ci}+1)}{\q_c(\q_c+1)} 
\end{align*} 
and
\begin{align*} 
\Var[p_{i|c}^2] &= \E[p_{i|c}^4] - \E[p_{i|c}^2]^2 \\
&= \frac{\q_{ci}(\q_{ci}+1)(\q_{ci}+2)(\q_{ci}+3)}{\q_c(\q_c+1)(\q_c+2)(\q_c+3)} - \left(\frac{\q_{ci}(\q_{ci}+1)}{\q_c(\q_c+1)}\right)^2
\end{align*} 
The last term consists of 
\begin{align*} 
\sum_{c \neq c'} \E[p_{i|c}^2]\E[p_{i|c'}^2] \Cov[p_c,p_{c'}] &= -\sum_{c \neq c'}\frac{\q_{ci}(\q_{ci}+1)}{\q_c(\q_c+1)} \frac{\q_{c'i}(\q_{c'i}+1)}{\q_{c'}(\q_{c'}+1)} \frac{\q_c\q_{c'}}{\q^2(\q+1)}.
\end{align*} 

For larger data sets, the following approximation can be made to keep things tractable and computationally feasible. We approximate the variance of $p_{ij}$ by assuming that there is no uncertainty about $p_c$, and assume $p_c = \p_c$. We then have 
\begin{align*} 
\Var[p_{ij}] &\approx \sum_c \p_c^2 \Var[p_{i|c}p_{j|c}].
\end{align*} 
This gives for $i \neq j$:
\begin{align*} 
\Var[p_{ij}] &= \sum_c \p_c^2 (\E[p_{i|c}^2p_{j|c}^2] - \E[p_{i|c}p_{j|c}]^2)\\
&= \sum_c \frac{\q_c^2}{\q^2} \left(\frac{\q_{ci}\q_{cj}(\q_{ci}+1)(\q_{cj}+1)}{\q_c(\q_c+1)(\q_c+2)(\q_c+3)} - \frac{\q_{ci}^2\q_{cj}^2}{\q_c^2(\q_c+1)}\right)
\end{align*} 
and for $i=j$
\begin{align*} 
\Var[p_{ii}] &= \sum_c \p_c^2 (\E[p_{i|c}^4] - \E[p_{i|c}^2]^2)\\
&= \sum_c \frac{\q_c^2}{\q^2} \left(\frac{\q_{ci}(\q_{ci}+1)(\q_{ci}+2)(\q_{ci}+3)}{\q_c(\q_c+1)(\q_c+2)(\q_c+3)} - \left(\frac{\q_{ci}(\q_{ci}+1}{\q_c(\q_c+1)}\right)^2\right)
\end{align*}

\section{MCMC simulations for $PMI(p_{ci})$ and $PMI(p_{ij})$}\label{app:MCMC} 

\begin{figure*}[ht]
\centering
\includegraphics[width=\linewidth]{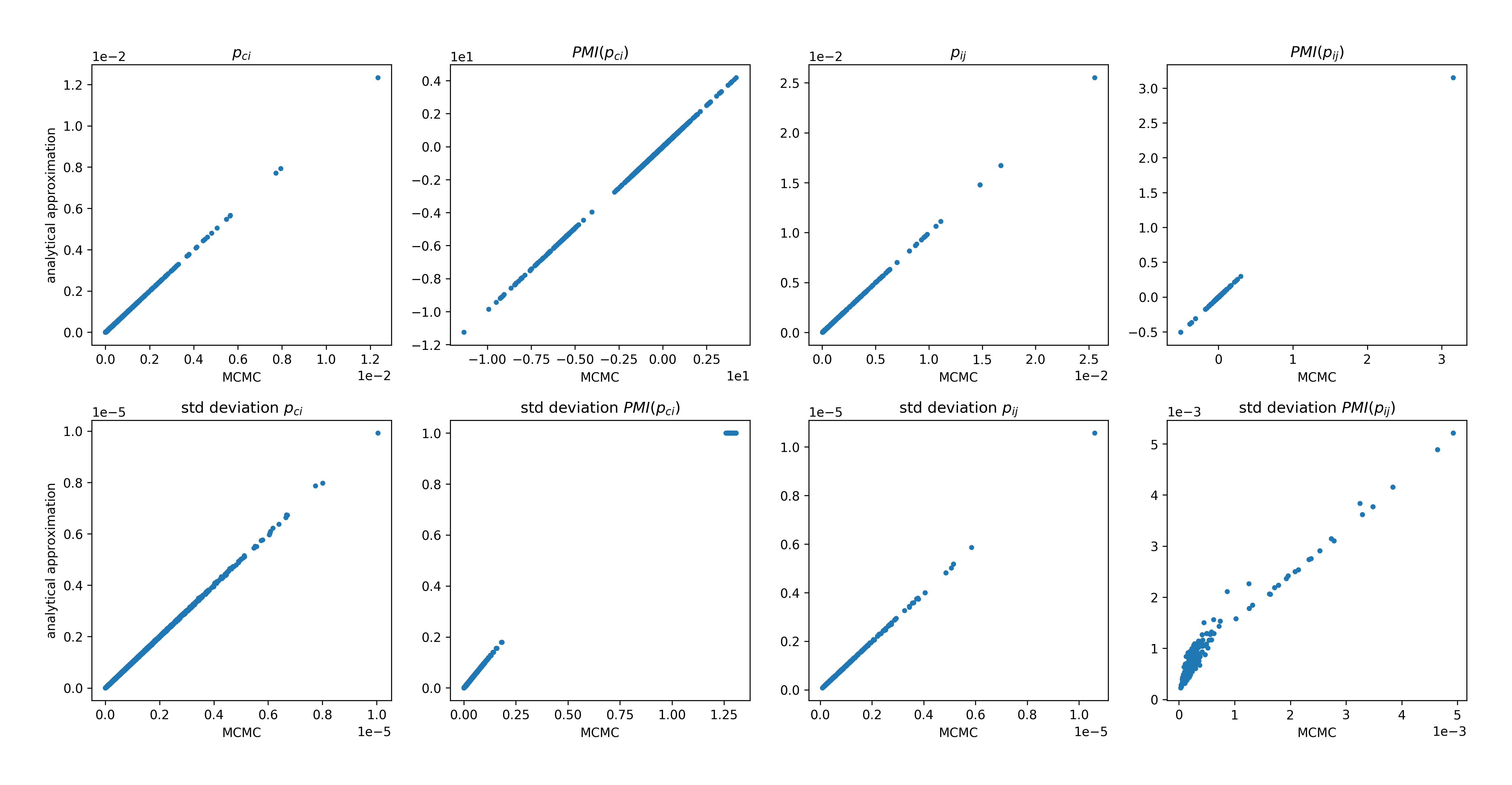}
\caption{Simulated values versus analytical approximations. }

\label{fig:MCMC}
\end{figure*}

We implement the estimation procedure in Python using the package pymc3. Figure \ref{fig:MCMC} compares the analytical approximations to Markov chain Monte Carlo simluations. The results show a good fit between simulated results and the analytical pproximations, except for the standard deviation of the $PMI(p_{ij})$, where the analytical approximation slightly over-estimates the standard deviation. 

\section{Estimation of $\Var[KL(p_{j|i}|p_j]$}\label{app:varKL}

We estimate $KL(p_{j|i}|p_j)$ in a similar way as the $PMI(p_{ij})$. By computing the Taylor expansion and taking we expectation we obtain
\begin{align*} 
\E[KL(p_{j|i}|p_j)] &\approx KL(\hat{p}_{j|i}|\hat{p}_j) + \sum_j \Var(p_{ij}) \frac{\partial^2}{\partial p_{ij}^2} p_{j|i} \log\left(\frac{p_{ij}}{p_i p_j}\right)\\
&= KL(\hat{p}_{j|i}|\hat{p}_j) + \sum_j \Var(p_{ij}) \left(\frac{2p_{ij}}{p_ip_j^2} + \frac{p_{ij}}{p_i^2p_j} + \frac{3p_{ij}}{p_j}^3 + \frac{1}{p_i p_{ij}} - \frac{4}{p_j^2} - \frac{2}{p_ip_j} + \left(\frac{2p_{ij}}{p_j^3} - \frac{2}{p_j^2}\right) \log\left(\frac{p_{ij}}{p_ip_j}\right)\right).
\end{align*} 

For the variance of $KL(p_{j|i}|p_j)$ we obtain, using the delta method,
\begin{align*} 
\Var(KL(p_{j|i}|p_j) &\approx \sum_j \Var(p_{ij}) \frac{\partial}{\partial p_{ij}}  p_{j|i} \log\left(\frac{p_{ij}}{p_i p_j}\right)\\
&= \Var(p_{ij}) \left(\frac{1}{p_j} + \frac{p_{ij}}{p_j^2}\right)\left(\log\left(\frac{p_{ij}}{p_ip_j}- 1\right) - \frac{p_{ij}}{p_ip_j}\right). 
\end{align*} 
The standard deviations are given as the square root of this variance. Similar equations can be derived for $KL(p_{c|i}|p_c)$ and $KL(p_{i|c}|p_i)$.

\end{document}